\documentclass{appolb}
\usepackage{amssymb,amsmath}
\usepackage{epsfig}

\def\Journal#1#2#3#4{{#1} {\bf #2}, #3 (#4)}

\def\PRpC{{\em Phys. Rep.} C}
\def\PLB{{\em Phys. Lett.} B}
\def\ZPC{{\em Z. Phys.} C}
\def\CPC{{\em Comp. Phys. Comm.} }
\def\JETPL{{\em  JETP Lett. }}
\def\PRp{{\em Phys. Rep.} } 
\def\NPB{{\em Nucl. Phys.} B}
\def\PRD{{\em Phys. Rev.} D}
\def\MPLA{{\em Mod. Phys. Lett.} A}
\def\epem{$e^+e^-$}
\def\ednd3p{E\frac{dn}{d^3p}}
\def\edsd3p{E\frac{d\sigma}{d^3p}}
\newcommand{\N}{{\mathcal N}}
\newcommand{\Nq}{{\mathcal N}_q}   
\newcommand{\Ng}{{\mathcal N}_g}
\newcommand{\Na}{{\mathcal N}_a}
\newcommand{\Nb}{{\mathcal N}_b}
\newcommand{\Nc}{{\mathcal N}_c}

\def\lapproxeq{\lower .7ex\hbox{$\;\stackrel{\textstyle
<}{\sim}\;$}}
\def\gapproxeq{\lower .7ex\hbox{$\;\stackrel{\textstyle
>}{\sim}\;$}}

\begin{document}
\pagestyle{plain}
\newcount\eLiNe\eLiNe=\inputlineno\advance\eLiNe by -1
\title{LIMITING SOFT PARTICLE PRODUCTION AND QCD\thanks{Presented at the 50th Cracow School of Theoretical Physics, 9-19
June 2010 in Zakopane, Poland} 
}
\author{Wolfgang Ochs
\address{Max-Planck-Institut f\"ur Physik, F\"ohringer Ring 6, D-80805
Munich, Germany}}
\maketitle

\begin{abstract}
We present some basic elements of the treatment of particle multiplicities
in jets from high
energy collisions within perturbative QCD. Then we 
discuss the universal features of the inclusive particle spectrum 
for the limiting
case of momentum $p\to 0$ (or $p_T\to 0$) as expected from soft QCD gluon
bremsstrahlung. The energy independence
of the invariant particle density in this limit $I_0=E\frac{dN}{d^3p}|_{p\to 0}$ 
is predicted as well as the 
dependence of this quantity on the the colour factors characteristic of the
underlying partonic processes. These properties are
first recalled from $e^+e^-$ collisions and then extended to $pp$ and nuclear
collisions according to Ref. \cite{Ochs:2010sv}. 
Present data support these predictions.  It will be interesting to see
whether new incoherent contributions show up in the new energy regime of
LHC.  \end{abstract}
PACS numbers: 12.38 Bx, 12.38 Qk, 13.85 Hd
\section{Introduction}
The production of multi-hadron final states at high energies is 
described within QCD as a two-phase process: in the first
phase there is some hard scattering (strong or electroweak) 
of the incoming elementary objects like quarks,
 leptons or gauge bosons. The produced quarks and gluons (``partons'') 
will form jets
of partons by gluon bremsstrahlung and quark pair production
according to the rules of 
QCD perturbation
theory for a characteristic cut-off scale $Q_0$. In a second phase the
partons reinteract and hadrons are formed which ultimately 
decay into stable
particles. These processes are not accessible in perturbation theory and 
particular models are applied for their description.

The simplest and best understood high energy process 
is $e^+e^-$ annihilation into
hadrons. It is initiated by the process $e^+e^-\to q\bar q$  which evolves
into two  hadronic jets dominantly. In $pp$ collisions the protons in the
primary hard collision act as a collection of partons and in an event
triggered for large transverse energy the partons
scattered into large angles form sidewise jets while the spectator jets follow the direction of
the incoming protons. The soft particles in this case form the ``underlying
event'' and this phenomenon is under intense investigation today. In addition,
there is the possibility of multiple independent parton-parton interactions 
considered important at the highest available energies. There are
also the untriggered ``minimum bias'' events which may result from small
angle parton-parton scatterings. Finally, in nucleus nucleus collisions
there may occur hard parton parton interactions as in $pp$ collisions, in
addition there are parton reinteractions in the large nucleus, 
multiple nucleon-nucleon interactions and, with special
interest, the new collective phenomena like quark gluon plasma formation.

Although  very different phenomena appear in the various processes there are
some remarkable simplifications for very soft particles with momentum in the
limit $p\to0$ in all processes. 
We consider for inclusive particle distributions the limit
for the particle density
\begin{equation}
 I_0=\left. E\frac{dN}{d^3p}\right|_{p\to 0}. \label{i0}
\end{equation}
In this limit the Born term in the perturbative expansion dominates
and this leads to some universal features for all processes
\begin{enumerate}
\item inclusive spectra become energy independent
\item the relative normalization of spectra in different processes is given
by the colour factors relevant for the minimal partonic process.
\end{enumerate}
This holds for QCD partons, but we assume the same is true also for hadrons.

These properties can be  understood qualitatively as follows.
A soft gluon is coherently emitted from all final partons. Having a 
large wavelength it cannot resolve any detailed intrinsic jet structure.
It ``sees'' only the total colour charge which is carried by 
the primary partons, and these are represented by the Born term for the
minimal partonic process in the perturbative expansion.

\section{Inclusive properties of QCD jets}
\subsection{QCD evolution equations}
We begin by recalling the main tools to derive the
inclusive observables for parton jets. They are obtained 
analytically in QCD
using the concept of evolution equations (see, for example, Refs.
\cite{dkmt:1991,Khoze:1996dn}, some more recent results will be added). 
Let us consider the partons in a jet emerging
from a primary parton of energy $E$ within the opening angle $\Theta$. 
First we
consider the global observables like mean multiplicity $\langle n\rangle$,
factorial moments $f_q=\langle{n(n-1)\ldots (n-q+1)}\rangle$ of the
multiplicity distribution which can be derived
from a generating function 
\begin{equation}
Z(Q,u)=\sum_{n=1}^\infty P_n(Q) u^n
\end{equation}
for the jet scale $Q=E\Theta$ at small angles $\Theta$ 
and the probability $P_n$ for production of $n$ particles as
\begin{equation}
\bar n= \left. \frac{\partial Z(Q,u)}{\partial u}\right|_{u=1},\quad
f_q=\left. \frac{\partial^n Z(Q,u)}{\partial u}\right|_{u=1}.
\label{nbarfq}
\end{equation}
In a corresponding way we obtain inclusive distributions 
$D(k)\equiv \frac{dn}{d^3k}$, i.e. the number of particles in the interval
$d^3k$, and, more generally, inclusive correlation functions
 $D^{(n)}(k_1\ldots k_n)$ from a generating functional which depends on a
probing function $u(k)$
\begin{gather}
Z(Q,{u})=\sum_n \int d^3k_1\ldots d^3k_n P_n(k_1\ldots k_n)u(k_1)\ldots
u(k_n)\\
D^{(n)}(k_1\ldots k_n)=\left. \frac{\delta^n Z(\{u\})}{\delta(u(k_1)\ldots
\delta(u(k_n)}\right|_{u=1}
\end{gather}
where $P_n$ is the probability distribution of momenta $k_i$.

For this generating function or functional an evolution
equation is derived in the scale $Q=E\Theta$ 
which yields, by appropriate differentiation, the equations
for the observables like multiplicities and inclusive spectra. In
differential form one finds the coupled equation for quark and gluon jets
($a=q,g$)  \cite{dt,dkmt:1991}.\footnote{Simplified forms have been
obtained before. \cite{dfk1,bcm}} They can be considered as  extensions of
the well known
DGLAP evolution equations  
towards low particle energies
 taking into account 
soft gluon coherence as realized in a probabilistic way by angular ordering
\cite{ao,ahm2}:
\begin{eqnarray}
\frac{d}{dY} Z_a(Y,u)&=& \sum_{b,c}\int_{z_c}^{1-z_c} dz
   \frac{\alpha_s(\tilde{k_T})}{2\pi}P_{bc}(z)\times \nonumber\\  
     && \hspace{0.2cm} \{Z_b(Y + \ln z,u)Z_c(Y + \ln (1-z),u) - Z_a(Y,u)\}
\label{Zevol}\\
Z_a(0,u)&=& u \label{Zinit} 
\end{eqnarray}
The evolution variable is taken as $Y=\ln ({E\Theta}/{Q_0})$ with the
non-perturbative $k_T$ cut-off $Q_0$; the argument of the running coupling 
is  $ \tilde k_T = \min (z,1-z)E\Theta$. The evolution equation
(\ref{Zevol})
describes the decay of a parton jet $a$ at scale $E$ into two parton jets 
$b,c$ at scale
$zE$ and $(1-z)E$ with probability $P_{bc}(z)$, the so-called DGLAP splitting
functions. The second equation (\ref{Zinit}) represents the initial
conditions at threshold ($E\Theta=Q_0$) and means that the parton jets start
just with the initial parton $a$.  
  
Asymptotic solutions can be obtained in the Double Logarithmic Approximation
(DLA) which includes only the dominant contributions from the singularities
at small angles and energies in the emission probability.
In this approximation the splitting function 
$P_{gg}(z)\sim 1/z$ in (\ref{Zevol}); 
the next to leading single logarithmic 
terms are included in the Modified Leading Logarithmic Approximation (MLLA). 
Up to this order the results from Eq. (\ref{Zevol}) are complete;
further logarithmic contributions beyond MLLA can be calculated, but they
are
not complete and neglect in particular process dependent large angle
emissions.
 Nevertheless they improve the results considerably as they take into
account  
energy conservation with increasing accuracy. The full 
solution of Eq. (\ref{Zevol}), corresponding to the summation of all
logarithmic orders can
be obtained numerically. Alternatively, one may calculate results of the QCD   
cascade from a Monte Carlo generator, such as ARIADNE  
\cite{ARIADNE}, which applies the same $k_T$ cut-off procedure as
Eq. (\ref{Zevol}).  

\subsection{Parton Hadron Duality Approaches}
So far we have discussed the properties of a jet of partons obtained from
perturbation theory using an artificial cut-off at low scales $Q_0$.
The application to multiparticle observables needs an additional assumption
about the hadronization process
at large distances which is governed by the color-confinement
forces not accessible by perturbation theory.

The simplest idea is to treat hadronization as long-distance
process, involving only small momentum transfers, and to compare
directly the perturbative predictions at the partonic level with the
corresponding measurements at the hadronic level. This can be
applied at first to the total cross sections, where at the low energies the
resonance structures are represented in an average sense. The perturbative
approach also describes jet
production for a given resolution; here the collection of partons is compared
to hadronic jets at the same resolution and kinematics. This approach has
led to spectacular successes and has built up our present confidence in
the correctness of QCD as the theory of strong interactions. 

In a next step one may carry on such a dual correspondence
further to the level of partons and final hadrons.
This procedure turns out successful for ``infrared and
collinear safe'' observables which do not change if a soft
particle is added or one particle splits into two collinear particles.
Such observables become insensitive to the cut-off $Q_0$ for small $Q_0$.
Quantities of this type are energy-flows and -correlations as well as
global event shapes like thrust etc.

Further on, one may compare partons and hadrons for
observables which count individual particles, for example, particle
multiplicities, inclusive spectra and multiparton correlations.
Such observables depend explicitly on the cut-off $Q_0$ (the smaller
the cut-off, the larger the particle multiplicity).

According to the hypothesis of 
Local Parton Hadron Duality (LPHD)~\cite{Azimov:1984np} 
the hadron spectra are proportional to the parton spectra
where the conversion of partons into hadrons occurs at a 
low virtuality scale, of the order of hadronic masses, i.e. $Q_0\sim $ few
hundred MeV,
independent of the scale of the primary hard process.

While this hypothesis has been suggested originally for single inclusive
spectra it can be generalized to more complex situations of the form
\cite{Khoze:1996dn}
\begin{equation}
 O(x_1,x_2,\ldots)|_{hadrons}= K~O(x_1,x_2,\ldots, Q_0, \Lambda)|_{partons},
\label{lphdeq}
\end{equation}
where the non-perturbative cut-off $Q_0$ and the ``conversion coefficient''
$K$ have to be determined by experiment. 
The conversion coefficient should be a true
constant independent of the hardness of the underlying process.

In a more recent analysis 
mean multiplicities and higher multiplicity moments have been calculated
both for sub-jets
of variable cut-off scale $Q_c$ (``jet virtuality'') and for hadrons with cut-off
$Q_0$ in $e^+e^-$ annihilation \cite{Lupia:1997in,Buican:2003jx} with the
smooth transition from jets to hadrons for $Q_c\to Q_0$.\footnote{For jets
and sub-jets the so-called Durham-algorithm \cite{durham} which
corresponds to a cut-off $k_T>Q_c$ has been applied.} For jets at fixed cut-off
$Q_c$ the normalisation is $K=1$ in (\ref{lphdeq}). 
It turns out that a unified description of jets and hadrons 
was possible with the common
normalization
\begin{equation}
K\approx 1. \label{keq1}
\end{equation}
In this case the hadronic cascade has been represented by the partonic
cascade in the average with the same multiplicity of partons and hadrons.  
So the
resonance bumps in the multi-particle spectra are just represented by 
the corresponding smooth perturbative spectra in the average. This
{\it parton-hadron-jet correspondence} implies that 
{\it a hadron corresponds to a parton jet of 
resolution $Q_0$}.

When comparing differential
parton and hadron distributions there
can be a mismatch near the soft limit caused by the
mass effects (partons are taken as massless in general). 
This mismatch can be avoided by a proper choice
of energy and momentum variables. In a simple
model$\,$\cite{Lupia:1995hb,{Khoze:1996ij}}
partons and hadrons are compared at the same energy (or transverse mass)
using an effective mass $Q_0$ for the hadrons, i.e.
\begin{equation}
E_{T,parton}=k_{T,parton}\Leftrightarrow E_{T,hadron}=
\sqrt{k^2_{T,hadron} + Q^2_0},
\label{partonhadron}
\end{equation}
then, the corresponding lower limits are $k_{T,parton}\to Q_0$
and $k_{T,hadron}\to 0$.

We should remark that these duality approaches are justified primarly by their
phenomenological success and their intrinsic simplicity 
and not yet by a convincing theoretical derivation from QCD. In particular
they allow compact analytical solutions for the observable 
quantities in the available
approximations to QCD (DLA, MLLA\ldots), which is not possible for the
phenomenological hadronization models of high complexity.   

\section{Quark and gluon jets: global observables vs. soft limit}
\subsection{Global observable: particle multiplicities}
Before we derive the soft properties of particle spectra we discuss 
the mean particle multiplicity in a jet as the most simple example of a global
event characteristic. Here the higher orders in the QCD perturbation theory
are very important. By differentiation (\ref{nbarfq}) of 
the evolution equations 
of the generating functions (\ref{Zevol}) one obtaines the evolution
equations for the parton multiplicities $\Na$ in quark and gluon jets ($a=q,g$)
\cite{Lupia:1997in}
\begin{eqnarray}
\frac{d \Na(Y)}{d Y}  & = &
\frac{1}{2}   
 \sum_{b,c} \int_0^1  dz 
\frac{\alpha_s (\tilde k_T)}{\pi}  P_a^{bc} (z) \nonumber \\
 \; &\times &
\left [\Nb (Y+\ln z) + \Nc (Y+\ln (1 - z))  -  
\Na (Y) \right ]  \label{eveqmult}
\end{eqnarray} 
with initial conditions
\begin{equation}
\Na (Y)|_{Y=0}  = 1,
\label{initmult}
\end{equation} 
which imply there is only one particle in a jet at threshold. Starting with
this initial condition one can obtain by iteration of the evolution equation
the perturbative expansion. 

The asymptotic behaviour can be derived from (\ref{eveqmult}) using the
ansatz
\begin{equation}
\N_g(Y) \sim  \exp\left(\int^Y\gamma(y)dy\right), \label{nasy}
\end{equation} 
where the anomalous dimension $\gamma$ has an expansion in 
$\gamma_0=\sqrt{2N_C\alpha_s/\pi}$
\begin{equation}
\gamma=\gamma_0(1-a_1 \gamma_0 - a_2 \gamma_0^2 - a_3 \gamma_0^3 \ldots),
 \label{gamma}
\end{equation}
and for the ratio of gluon and quark jet multiplicities
\begin{equation}
r\equiv \frac{\Ng}{\Nq}=\frac{C_A}{C_F}(1-r_1\gamma_0 -
r_2\gamma_0^2-r_3 \gamma_0^3 \ldots)
\label{rgq}
\end{equation}
with QCD colour factors 
\begin{equation}
C_A=N_C=3, \quad C_F=\frac{4}{3}.
\end{equation} 	
The coefficients $a_i$
and $r_i$ can be derived from the evolution equations.
At high energies the leading behaviour in MLLA for both quark and gluon jet 
multiplicities is given by
\begin{gather}
\ln  \N(Y) \sim  c_1/\sqrt{\alpha_s (Y)} +  c_2\ln \alpha_s (Y) +c_3
\label{lnn2}\\
c_1  =  \sqrt{96\pi}/b,  \qquad   c_2=\frac{1}{4}+\frac{10}{27}n_f/b,
\qquad b=\frac{11}{3}C_A-\frac{2}{3}n_f.
\label{lnncoef}
\end{gather}
with the arbitrary constant $c_3$, and this behaviour describes well the
data in $e^+e^-$ annihilation at LEP-1 and LEP-2, for review, see 
\cite{dg}, more recent results have been presented by  DELPHI
\cite{Abreu:1999rs,Abdallah:2005cy,Siebel:2006np} and OPAL
\cite{Abbiendi:2001us,Abbiendi:2003gh}.

\begin{figure}[t!] 
\vspace*{8.5cm}
\begin{center}
\includegraphics{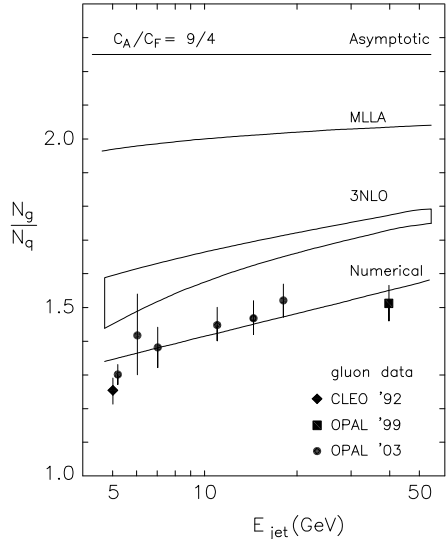}
\vspace{-1.0cm}
\caption[*]{ 
The ratio of the mean multiplicities in gluon jets and quark jets 
$N_g$ and $N_q$ obtained from $e^+e^-$ experiments;
results from perturbative QCD show the large higher
order corrections for a global observable (from \cite{Ochs:2005vt}). 
\vspace{-1cm}}
\end{center}
\label{fig:rgq}
\end{figure}

The important role of higher logarithmic orders can be studied in the behaviour 
of the multiplicity ratio $r$ in (\ref{rgq}). The asymptotic limit $r=C_A/C_F$
acquires large finite energy corrections
 in  NLLO
\cite{ahm1,mw} and 2NLLO order \cite{gm,dreminosc}
\begin{eqnarray}
 r_1 &=& 2\left(h_1+\frac{N_f}{12N_C^3}\right) -\frac{3}{4}\\
 r_2 & =& \frac{r_1}{6} \left(\frac{25}{8}- \frac{3N_f}{4N_C} -
                               \frac{C_FN_f}{N_C^2}
           -\frac{7}{8}-h_2-\frac{C_F}{N_C}h_3+\frac{N_f}{12N_C} h_4\right)
\end{eqnarray}
with $h_1=\frac{11}{24},\ h_2=\frac{67- 6\pi^2}{36},\ 
h_3=\frac{4\pi^2-15}{24}$ and $ h_4=\frac{13}{3}$, also 3NLLO results have
been derived \cite{cdgnt}. Results from these approximations \cite{dg}
are shown in Fig.~1 
together with the numerical solution of the MLLA evolution equations 
(\ref{eveqmult}) obtained in 1998 \cite{Lupia:1997in},
 which takes into account all higher order corrections from this equation
and fulfils the (non-perturbative) boundary condition (\ref{initmult}).
All curves are absolute predictions, as the parameter $\Lambda$ (and $Q_0$ in
case of the numerical calculation)
is adjusted from the growth of the total particle multiplicity
in the $e^+e^-$ jets.
The slow convergence of this $\sqrt{\alpha_s}$ expansion can be seen
and there are still considerable effects beyond 3NLLO. The 
numerical solution is also in close agreement with 
the MC result at the parton level obtained \cite{Abbiendi:2003gh} from the HERWIG MC 
above the jet energy 
$E_{\rm jet}>15$ GeV ($E_{\rm jet}=Q/2$ in $e^+e^-$ annihilation)
 and $\sim$ 20\% larger
at $E_{\rm jet}\sim 5$ GeV.
This overall agreement suggests that the 
effects not included in the MLLA evolution equation, 
such as large angle emission, are small.

These numerical results are also compared in Fig.~1 
with data from OPAL \cite{Abbiendi:2003gh} 
where the data on gluon jets are derived from 
3-jet events in $e^+e^-$-annihilation. Note also that a proportionality 
constant $K$
relating partons and hadrons
according to LPHD drops in the ratio $r$.
The results obtained from DELPHI 
\cite{Abdallah:2005cy} fall 
slightly below the curve by about 20\% at the lowest
 energies but converge for the higher ones; 
the CDF collaboration comparing quark and gluon jets at high $p_T$
in $pp$ collisions \cite{pronko} finds the ratio $r$ in the range 
$5<E_{\rm jet}<15$ GeV a bit larger, closer to the 3NLLO prediction,
 but with larger errors 
and therefore still consistent with the LEP results.

\subsection{Inclusive energy spectrum: soft limit}
Next, we consider the inclusive distribution  $D(\xi,Y)$ of partons in the momentum fraction
$x=k/E$ or $\xi=\ln(1/x)$
within a jet with primary parton energy $E$ and opening angle 
$\Theta$. The
evolution equation for $D$ can be obtained by functional differentiation of
(\ref{Zevol}) (for a review, see \cite{Khoze:1996dn}b). At small $x$ (large
$\xi$) the angular ordering \cite{ao} of the cascade evolution 
which takes into account the soft
gluon interferences in a probabilistic way plays an important role. For
large $x$ the equations approach the well known DGLAP evolution equations. 

For the present discussion we 
restrict ourselves to the simplest approximation, the DLA with fixed
coupling, where only the most singular terms in the splitting functions for
$g\to gg$ and $q\to qg$ are kept. Then the evolution equation for parton $a$ 
reads
\begin{equation}
D_a^g(\xi,Y)=\delta_a^g\delta(\xi) +\int_0^\xi d\xi'\int_0^{Y-\xi}
dy\frac{C_a}{N_C}\gamma_0^2(y) D_g^g(\xi',y). \label{Devol}
\end{equation}
This equation can be solved by iteration. For fixed $\gamma_0\sim
\sqrt{\alpha_s}$ one obtains  the perturbative expansion
\begin{eqnarray}
D_a^g(\xi,Y)&=&\delta_a^g\delta(\xi)+ \frac{C_a}{N_C} \gamma_0^2 (Y-\xi)
+\frac{1}{2}\frac{C_a}{N_C} \gamma_0^4 \xi (Y-\xi)^2 + \dots \label{xidist}\\
      &=&\delta_a^g\delta(\xi)+ \frac{C_a}{N_C} \gamma_0 
        \sqrt{\frac{Y-\xi}{\xi}}\, I_1(2\gamma_0\sqrt{\xi(Y-\xi)})
\end{eqnarray}
with the modified Bessel function $I_1$. The coherent soft gluon emission 
leads to a depletion of the spectrum at large $\xi$, also called ``the
hump-backed plateau''. From the inclusive distribution (\ref{xidist}) one can
obtain the double differential distribution in energy and angle by the
differentiation over $Y$ which yields 
$\frac{dN_a}{d\xi dY}=\frac{C_a}{N_C} \gamma_0^2+\frac{C_a}{N_C} \gamma_0^4
\xi (Y-\xi) + \dots$, or, in the
original variables
\begin{equation}
 \displaystyle  
\frac{dN_a}{dk d\Theta} = \frac{2}{\pi}\frac{C_a}{k \Theta}\alpha_s
 + \frac{4N_C}{\pi^2}\frac{C_a}{k \Theta}\alpha_s^2  
    \ln\frac{E}{k}\ln \frac{k_T}{Q_0}
+\ldots \label{Dborn}
\end{equation}
Here we recognize in the leading term of $O(\alpha_s)$ the well known Born
term for soft gluon bremsstrahlung as in QED, but with the appropriate QCD
colour factors. One observes that only the Born term survives in
the soft limit where $k_T\to Q_0$; in this limit we find the simple
universal properties emphasised in the introduction: 
the particle density becomes
independent of energy $E$ and is proportional to the relevant colour factor
for the minimal process, that is here the 
gluon emission from the quark or gluon jet with  $C_a=C_F$ or $C_a=C_A$
respectively.

\begin{figure}[t]
\epsfig{file=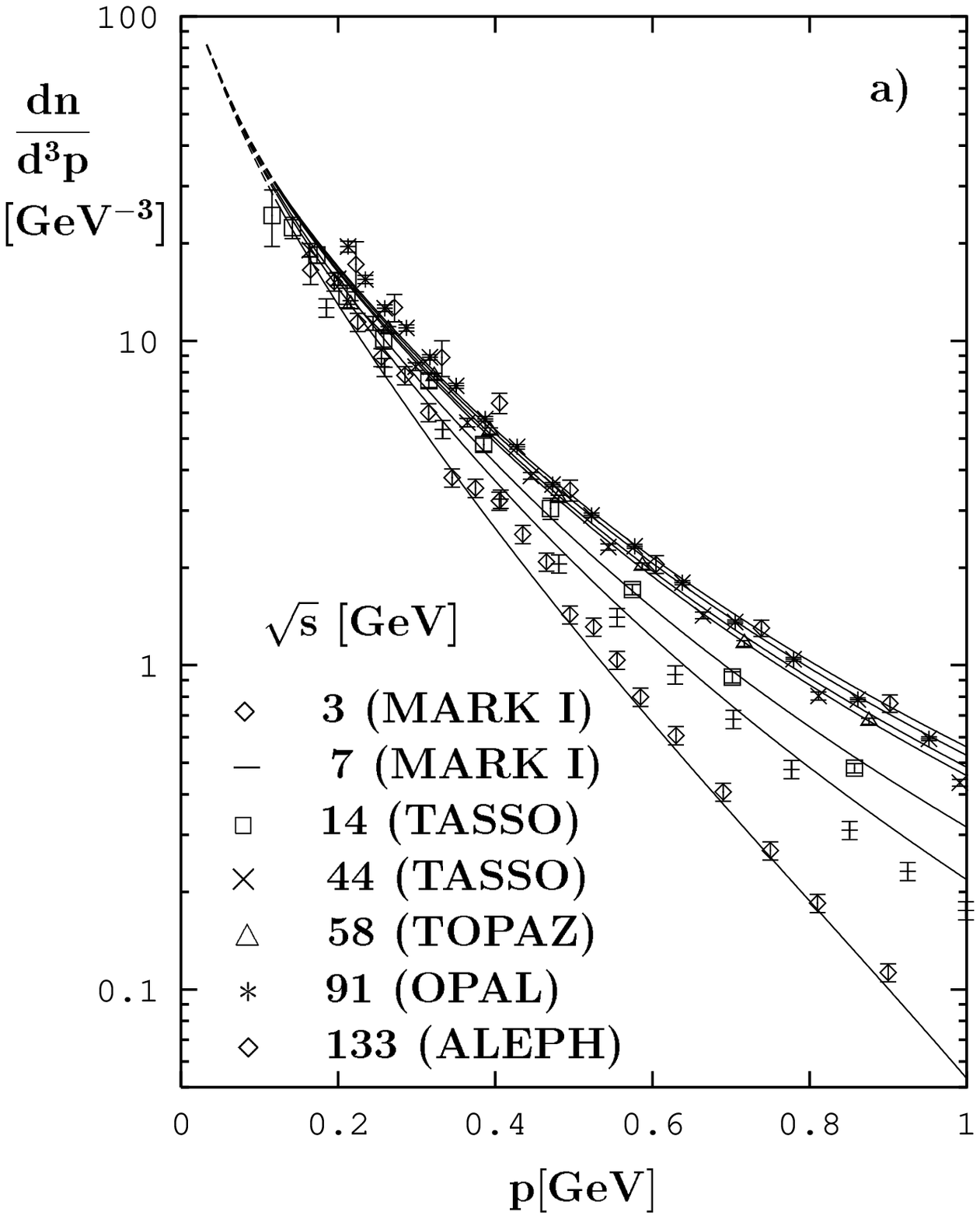,bbllx=4.5cm,bblly=9.5cm,%
bburx=16.5cm,bbury=26.cm,width=5.5cm}\hspace{1cm}
\epsfig{file=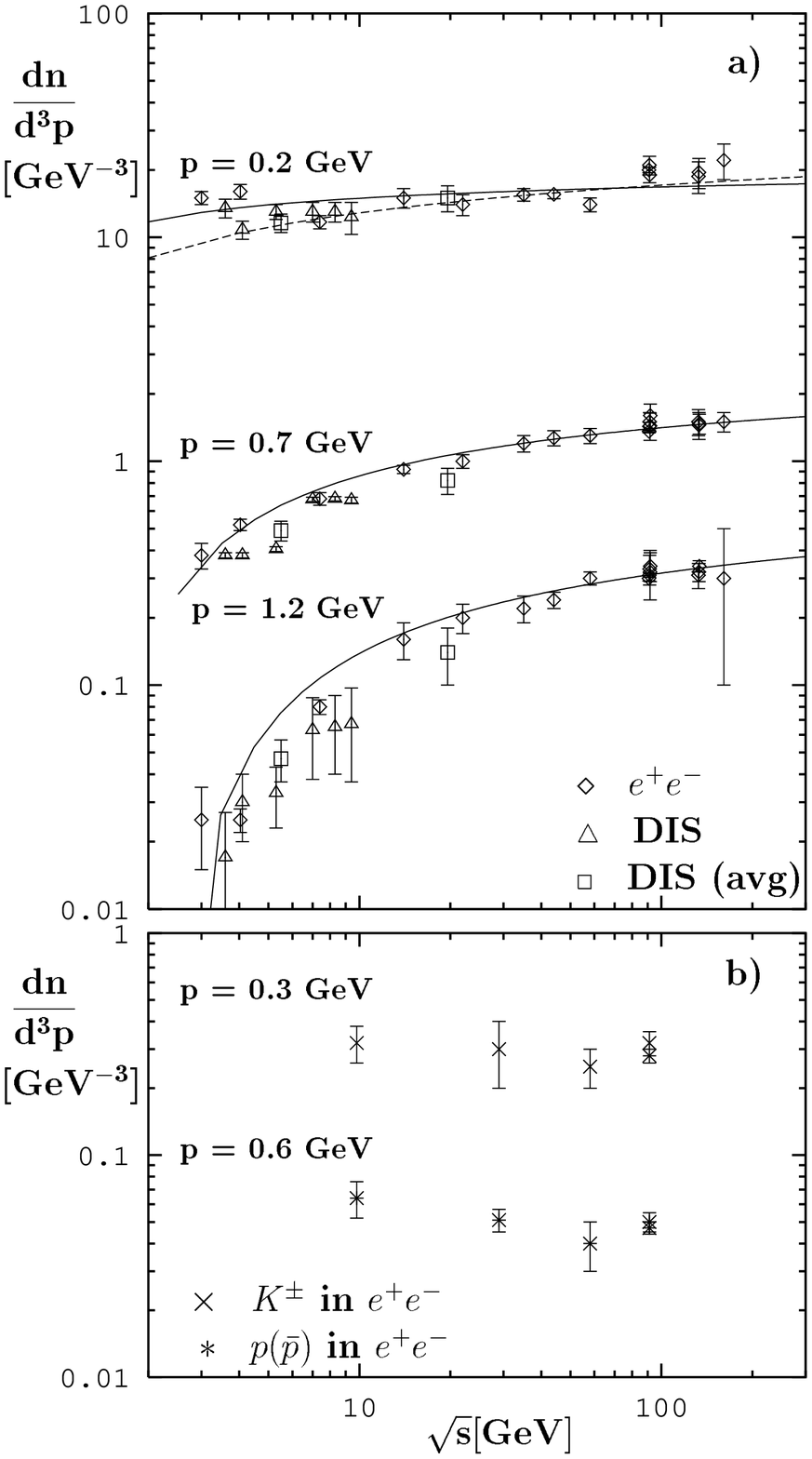,bbllx=4.5cm,%
bblly=4.5cm,bburx=16.5cm,bbury=26.cm,width=5.5cm}
\vspace{0.3cm}
\caption{Inclusive particle densities $\frac{dn}{d^3p}$ of charged particles 
in $e^+e^-$ annihilations; left panel: momentum
spectra for different energies, right panel energy $\sqrt{s}$-dependence for
fixed momenta $p$, lower right panel: kaons and protons; calculations 
within MLLA accuracy.}
\label{fig:p-spectra}
\end{figure}
This result can be generalised to the accuracy of DLA 
with running coupling which is
obtained by iterating (\ref{Devol}) accordingly up to $O(\alpha_s^2)$ 
which is appropriate for the low momentum region, furthermore results within
MLLA have been derived as well
\cite{Khoze:1996ij}. The above properties in the soft limit remain unaltered.
A comparison of these calculations is shown in Fig.
\ref{fig:p-spectra} where a model dependent kinematic relation as in 
(\ref{partonhadron}) is used. One can see that the data are rather well
described by the model which predicts an energy independent particle density
in the soft limit $p\to 0$.  

\subsection{Colour factors in quark and gluon jets}
In order to check the sensitivity to the colour factors in the Born term
we should study the dependence of the soft limit of momentum spectra in
quark and gluon jets. Separating quark and gluon jets results in
uncertainties increasing with smaller momenta. Therefore an alternative
procedure has been suggested \cite{Khoze:1996ij} which studies the radiation
into a cone perpendicular to the production plane of 3-jet events in 
\epem annihilation. 

One can consider two extreme limits with two jets aligned:\\
a) a quark and a gluon jet are parallel and recoil against a quark jet,
in this configuration the soft perpendicular radiation cannot separate the
two parallel jets and the intensity is as in a $q\bar q$ dipole,
proportional to the colour factor $C_F$;\\
b) the quark and antiquark are parallel and recoil against the gluon, 
in this case the primary configuration acts like a $gg$ dipole with colour
factor $C_A$.
Of course, in a realistic experiment one cannot go to such extreme limits
because of the finite width of the jets, but one can measure a certain
range of the angles in between the jets which interpolates between the
limits.

The soft gluon bremsstrahlung from a ``colour dipole antenna'' ($q\bar q$ or
$gg$) is given by \cite{dkmt:1991}

\begin{eqnarray}
 \displaystyle 
  \frac{dN_{A,F}}{d\Omega dk}&=&\frac{\alpha_s}{(2\pi)^2}\frac{1}{k}
W_{A,F}(\vec n_g) ,\quad W_{A,F}(\vec n_g)=2C_{A,F}(\widehat{i,j})\\
 \displaystyle (\widehat{i,j})&=&\frac{1-\cos
\Theta_{ij}}{(1-\cos\Theta_{is})(1-\cos \Theta_{js})},
\end{eqnarray}
where $\Theta_{ij}$ is the angle between partons $i$ and $j$ and $s$
denotes the soft gluon.
Then, for the aligned $q\bar q$ antenna one obtains
$W_F=4C_F/\sin^2\Theta_{qs}$.
 
In a 3-jet event $e^+e^-\to q\bar q g$ one finds in lowest order
\begin{equation}
W_{q\bar q g}(\vec n_g) =
C_A\ [(\widehat{q,g})+(\widehat{\bar q, g})-\frac{1}{N_C^2} 
(\widehat{q, \bar q})],   
\end{equation}
i.e. there are two dipoles between each of the quarks and the gluon of 
strength $C_A$
and a colour suppressed dipole between the $q$ and the $\bar q$. 
For $q\|g$ one finds $W=4C_F/\sin^2\Theta_{qs}$ like a $q\bar q$ dipole,
and for $q\|\bar q$ one obtains 
$W=4C_A/\sin^2\Theta_{gs}$ like a   $g\bar g$ dipole, as anticipated above.  

The radiation perpendicular to the 3-jet plane ($\cos \Theta_{is}=0$)
normalised to the same radiation in 2-jet events is then given by the simple
formula
\begin{eqnarray}
\displaystyle \frac{N_\perp^{q\bar q g}}{N_\perp^{q\bar q}}&=&
   \frac{W_\perp^{q\bar q g}}{W_\perp^{q\bar q}} \equiv \frac{C_A}{C_F}\
r(\Theta_{ij}) \label{3jetsofta}\\
 r(\Theta_{ij})&=& \frac{1}{4}[ (1-\cos\Theta_{qg})+
  (1-\cos\Theta_{\bar qg}) - \frac{1}{N_C^2}(1-\cos\Theta_{q\bar q})]
\label{3jetsoft} 
\end{eqnarray}

Such a measurement has been carried out by the DELPHI collaboration
\cite{Abdallah:2004uu,Siebel2003}. The above  
formulae should apply in the soft limit where the Born term of
$O(\alpha_s)$ dominates. The $p_T$ spectra in the cone perpendicular to the
3-jet plane are found all very similar for $p_T\lapproxeq 1$ GeV, 
therefore one can
study instead the integrated multiplicity in the respective cones.
One observes first that the multiplicities in the cone is well described by
the above formula (\ref{3jetsoft}), the data are accurate enough to even
notice the $1/N_C^2$ term in  (\ref{3jetsoft}). Furthermore, the data for
the ratio on the l.h.s. of (\ref{3jetsofta}) are found to depend linearly on
the function $r(\Theta_{ij})$ and one obtains from the slope
\begin{equation}
\frac{C_A}{C_F}=2.211\pm0.014(stat.)\pm 0.053(syst.)
\end{equation}
which is well consistent with the expected $C_A/C_F=9/4$ in QCD. 

From these studies of jets in \epem annihilation we can conclude that the
soft particle density indeed follows the prediction of the soft gluon Born
terms emphasized in the introduction\\
a) The spectra become independent of energy for $p\to 0$;\\
b) the soft particle density varies with the orientation of the colour
antenna as predicted, this implies that the soft particle density in quark and gluon jets
becomes proportional to the colour factors $C_A$ and $C_F$.
This is in strong contrast to the behaviour of global observables like
the mean multiplicity, which obtains large higher order corrections
from the integral over the perturbative expansion such as (\ref{Dborn}) 
and the ratio $r=N_g/N_q$ is as low as 1.5 at LEP energies instead of 2.25
(see Fig. 1).
In the soft limit there is no phase space for subsequent emissions, nor for
energy momentum conservation effects. 
With this experience we now investigate the hadronic collisions.  
 
\section{High energy $pp$ collisions}\
We discuss here the ``minimum bias'' events, which we consider as
non-diffractive events. In order to estimate the very soft particle
production we look for the minimal partonic process which
could be responsible for the soft gluon bremsstrahlung. We assume that
the relevant process 
is  a semihard $2\to 2+g_s$ scattering of lowest perturbative order 
where any two partons inside the proton
can scatter with one-gluon exchange at small angles. The exchange of a colour
octet gluon at small angle transfers the colour from the colour singlet
protons to the two outgoing
partonic clusters which are the colour octet sources of soft gluon 
bremsstrahlung. In the minimal configuration each proton splits into a
quark-diquark pair which scatter by gluon exchange. In large $N_C$
approximation the process corresponds to two radiating colour triplet 
antennae responsible for bremsstrahlung from initial and final partons.
It should be added that also more complex partonic processes will end up in
the production of two colour octet systems as discussed in~\cite{Ochs:2010sv}.

Therefore, according to our general rules, 
we expect the energy independent limiting soft radiation density 
in $pp$ collisions $I_0^{pp}$ for $p\to
0$ and, furthermore, we expect this density to occur in a fixed ratio to 
the corresponding
density in $e^+e^-$ collisions as ratio of colour octet and colour triplet
dipole sources
\begin{equation}
p\to 0:\qquad   I_0^{pp}/I_0^{e^+e^-}\approx C_A/C_F, \label{ratio94}
\end{equation}   
just like the ratio of the spectra in gluon and quark jets
discussed in the last section. 

This kind of relation (\ref{ratio94}) 
has been suggested by Brodsky and Gunion already in 1976 \cite{bg},
but relating the
integrated multiplicities in the central rapidity region to these colour
factors. From our QCD analysis we find Eq. (\ref{ratio94}) to be valid only in
the soft limit while the ratio of integrated multiplicities is found closer to unity (see
below). Similarly, relations of the kind (\ref{ratio94}) appear in some
early phenomenological models, but again for the integrated densities only, 
as outlined in  \cite{Ochs:2010sv}.

Our expectation of an energy independent $I_0^{pp}$ is based on a coherent
process.  It would be violated if there were multiple parton-parton
interactions (processes like $4\to 4+g_s$) added incoherently.  Such
processes appear in some models at high energies (see, for example, PYTHIA
\cite{TS1}) and so the measurements at LHC can shed some light onto the
contributions from such processes.

We have studied the energy dependence of $I_0^{pp}$ using the results of
fits to the invariant cross sections $\edsd3p$ measured in the energy range
$\sqrt{s}= 20\ldots 1800$ GeV obtained from the colliders at CERN, Fermilab
and Brookhaven and measurements of inelastic cross sections $\sigma_{in}$. The $p_T$ spectra look qualitatively similar to those in
Fig. \ref{fig:p-spectra} for \epem annihilations 
converging towards small $p_T$ but
falling more steeply at high momentum.
The $p_T$ spectra have been fitted to
distributions which at small $p_T$ behave like 
\begin{equation}
\edsd3p= A \exp(B p_T
+\ldots).\label{invfit}
\end{equation}
These fits are good down to the smallest measured $p_T\sim 0.1$
GeV. Then one finds  $I_0^{pp}=A/\sigma_{in}$ from extrapolation
$p_T\to 0$. The functional form (\ref{invfit}) is not analytic at $p_T=0$
and is therefore theoretically not satisfactory. This problem is avoided
using the ``thermal'' parametrisation in terms of $m_T$ instead of $p_T$
\begin{equation}
\edsd3p= \frac{A}{(\exp(m_T/T)-1)};\quad m_T=\sqrt{m^2+p_T^2} 
\label{invfitth}
\end{equation}
as applied by PHOBOS \cite{phobos} in nuclear collisions and a good fit
down to the smaller $p_T\sim 0.03$ GeV has been obtained. The extrapolated
values $I_0^{pp}$  are smaller by about 25\% as compared to the fit (\ref{invfit}).
 
\begin{figure}[t]
\begin{center}   
\mbox{\epsfig{file=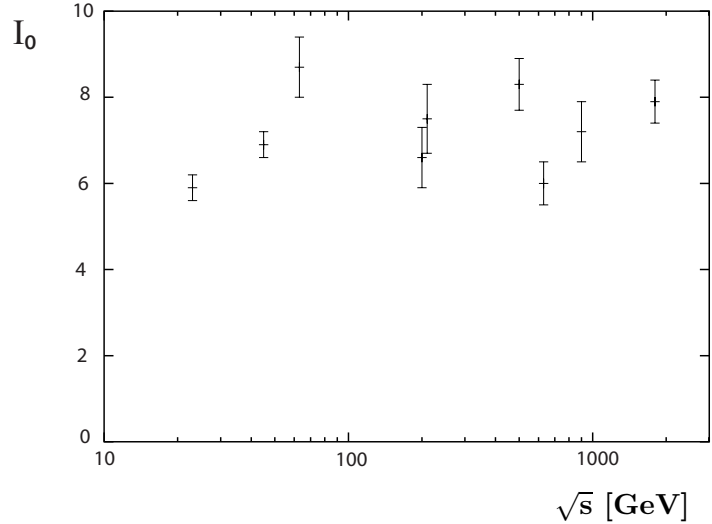,angle=0,bb=0 30 510 380,clip=,width=10cm}}
\end{center}
\vspace{-0.5cm}
\hspace*{9cm}{\bf ${\bf \sqrt{s}}$ [GeV] }
\caption{Soft limit $I_0^{pp}$ of the invariant density $E\frac{dn}{d^3p}$
of charged particles [$(h^+ +h^-)/2$] in $pp$ collisions as a function of
c.m.s. energy $\sqrt{s}$ (from exponential extrapolation).}
\label{fig:I0}
\end{figure}

The results from the available published 
exponential extrapolations are shown in Fig. \ref{fig:I0}. 
One observes a rather flat energy dependence over the two decades in energy
with an average $I^{pp}_0\approx (7\pm 1)$ GeV$^{-2}$. Note, that over this
energy range the rapidity density $\frac{dN}{dy}$ would rise by about a
factor 2. The different extrapolations (\ref{invfit}) or (\ref{invfitth})
should not affect the trend of the energy dependence. For comparison with
\epem annihilation it is better to normalise by 
the non-diffractive cross section which
we take as 15\% lower than the inelastic one, which yields
$I^{pp}_{0,nd}\approx (8\pm 1)$ GeV$^{-2}$. Then for the thermal fit 
and non-diffractive (minimum bias) events we obtain
\begin{equation}
I^{pp}_0\approx (6\pm 1)\ {\rm GeV}^{-2}.
\label{I0pp}
\end{equation}
So far, there is not yet a fit result from LHC to be used for comparison.

Next, we also compare this result with the soft limit in \epem annihilation
to test our prediction (\ref{ratio94}). We present two results.

1. There is a dedicated comparison by the TPC collaboration~\cite{tpc} 
who compared their own data on \epem annihilation 
 with $pp$ data from the British
Scandinavian collaboration \cite{Alper:1975jm} on the $p_T$ spectra of the
invariant density. The TPC data are presented
as function of $p_T$ as determined from the sphericity jet axis. 
In the average over $p_T$ both data sets for pions, 
kaons and protons are similar. A
closer look, however, reveals, that the spectra fall
more steeply with $p_T$ 
in the $pp$ collisions and there is a cross over of the spectra
at low $p_T$. The appropriate extrapolation down to $p_T$ near zero yields a
larger density for $pp$ collisions by a factor $2.0-2.7$ depending on the kind
of fit.

2. Most other experiments present fits to the spectra in particle energy
$E$ (not $p_T$). Using the fit results from various experiments in the range
$\sqrt{s}=10\ldots 29$ GeV yields 
$I_0^{e^+e^-}\approx (3.3\pm0.5)$ GeV$^{-2}$ or the ratio
\begin{equation}
I_0^{pp} / I_0^{e^+e^-}  \approx (1.8\pm 0.4) \div  (2.4\pm 0.5),
\text{          }\label{I0pe}
\end{equation} 
where the first (preferred) number refers to the thermal and the second to the
exponential extrapolation. This result is consistent with our QCD based
expectation for this ratio $C_A/C_F=2.25$.

\section{Nucleus-nucleus scattering}
For the nucleus-nucleus ($AA$) cross sections we may consider two 
limiting cases in the relation to the $pp$ cross section.

1. In case of a point like interaction the particle densities in 
nuclear collisions are obtained by rescaling the densities in $pp$
collisions by $N_{coll}$, the number of nucleon-nucleon collisions, or,
``the nuclear modification factor''
\begin{equation}
R_{AA}^{N_{coll}}=\frac{1}{N_{coll}}\frac{dN_{AA}/dp_T}{dN_{pp}/dp_T},
\label{RAB}
\end{equation}
is unity. The number $N_{coll}$ can be obtained from the Glauber model.

2. In case of soft particle production we expect that such particles with a
large wave length $1/p_T \gapproxeq r$ are coherently emitted over a range
$r$ (from a nucleon $r_N\sim 1/m_\pi$ or a nucleus $r_A\sim 1/(30$ MeV),
which results in a reduced rate. Indeed, the inspection of the RHIC data
\cite{phobos,star,phenix,brahms} shows the ratio $R_{AA}^{N_{coll}}$
falling below unity for small $p_T$. 
An alternative way presenting data in the soft region is the normalisation
to the number of ``participating nucleons'' 
\begin{equation}
R_{AA}^{N_{part}}=\frac{1}{(N_{part}/2)}\frac{dN_{AA}/dp_T}{dN_{pp}/dp_T}.
\label{RABpart}   
\end{equation}
This concept has been introduced already in 1976 by 
Bialas, Bleszynski and Czyz \cite{bialas},
who found the number of participating nucleons, called there 
``wounded nucleons'' as relevant scaling factor for soft production
($R_{AA}^{N_{part}}\approx 1$), i.e.
each interacting nucleon
should be counted only once and the rescatterings of the same nucleon be   
disregarded.

Again we consider the energy dependence and the normalization of the particle
production at central rapidity in the limit $p_T\to 0$. A detailed study
for various centralities (from peripheral to central $AA$ collisions) by
PHOBOS \cite{phobos} shows that the inclusive $p_T$ spectra in
normalisation (\ref{RAB}) or (\ref{RABpart}) approach about the same densities at
200 and at 62.4 GeV. This implies that the energy dependence is the same for
nuclei and protons, i.e. there is no sizable energy dependence.

Concerning the normalisation, the same data \cite{phobos} show that 
the quantity (\ref{RABpart}) approaches unity for all centralities (within
about 30\%) at the lower limit of $p_T\approx 200$ MeV. The STAR
collaboration \cite{star2} has measured this quantity with high precision
down to 500 MeV and an extrapolation to the ``participant scaling'' 
for $p_T=0$
is indicated. We combined  fits of PHOBOS low $p_T$ and STAR AuAu data
at 200 GeV using the thermal parametrisation and also the STAR $pp$ data
which yields the result
\begin{equation} 
I_0^{AA}/I_0^{pp}\approx 160\pm 17,
\label{AA/pp}
\end{equation}
which agrees
with the calculated $N_{part}/2=172\ (\pm15\%)$, and is therefore consistent
with
\begin{equation}
p_T\to 0: \qquad   R_{AA}^{N_{part}}\to 1\quad \text{and}\quad
I_0^{AuAu} \approx  \dfrac{N_{part}}{2}\ I_0^{pp}.
\label{RAApart}
\end{equation}  
This density for $AuAu$ collisions is about six times smaller than expected for 
an incoherent superposition of
collisions with $N_{coll}=1040$, where this number is 
obtained from Glauber model calculations.
It is remarkable, that the ``wounded nucleon'' model works to the precision
of about 10\%, the accuracy of measurements and theoretical calculations.
Note that this agreement is obtained only in the limit 
$p_T\to 0$ as can be seen from the
STAR data \cite{star2}; already for $p_T=0.5$  GeV the deviation from
``participant scaling'' amounts to about 50\%.
 
\begin{figure}[t]
\begin{center}   
\epsfig{file=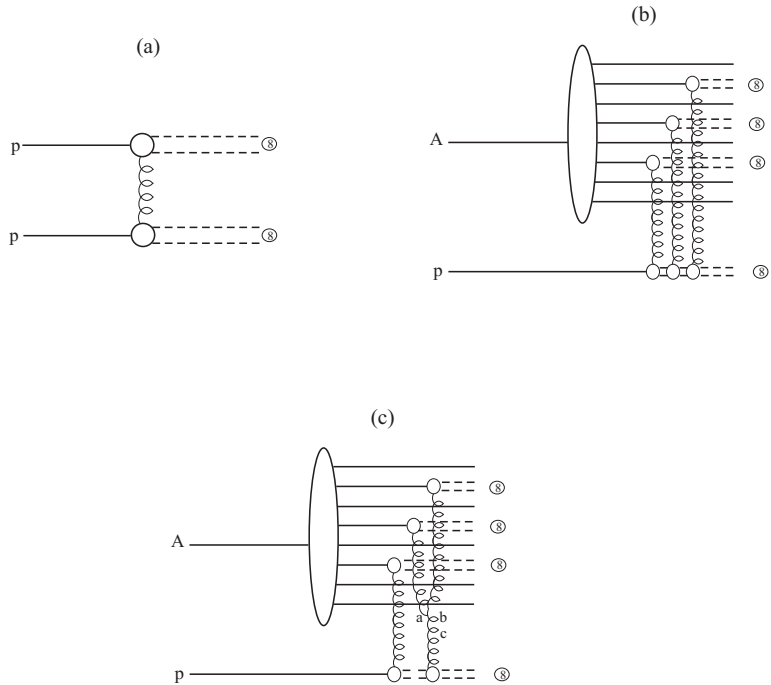,width=10.cm}
\end{center}
\caption{Diagrams, contributing to the $pp$ and $pA$ collisions in the
minimal model
for soft particle production: (a) In the $pp$ collisions the exchanged gluon  
interacts with the colour triplet constituents $q$ or $qq$ in the proton to
form an
outgoing colour octet system; (b,c) In the $pA$ collisions the proton can
rescatter
inside the nucleus and then forms a colour octet system again. This implies
that
the multiple gluon exchange acts as a single gluon exchange in the particle
production. In this example: $N_{coll}=3,\ N_{part}/2=2$.}
\label{fig:AA}
\end{figure}  

How can this scaling result be understood?  
An incoming nucleon scatters successively at a number of nucleons in the
nucleus (see Fig. \ref{fig:AA} b,c for representative diagrams of $pA$
scattering). The successive gluon exchanges yield again an outgoing colour
octet state as in $pp$ scattering, such that the rescatterings of the
nucleon are not causing any production of additional particles. This happens
if in the low $p_T$ interaction only a
quark and the diquark appear as active partons, so that also a multi-gluon
exchange cannot produce a higher colour multiplet than an octet.\footnote{A
model based on wounded quarks and diquarks has been developed in Ref.
\cite{bialas1}, but 
for the description of $p_T$-integrated rapidity distributions.} 
Alternatively, one can think of a larger number of exchanged gluons 
but taking into account that the colour octet exchange gives the 
dominant Leading Logarithmic contribution
both in DGLAP and BFKL kinematics. Then, from both viewpoints, 
each scattered nucleon 
produces dominantly a colour octet flow as in $pp$ interactions,
in agreement with the phenomenological result (\ref{RAApart}).

\section{Universal composition of softly produced particles}
\begin{figure}[t]
\begin{center}
\begin{minipage}[h]{0.5cm} 
\vspace{-4.5cm}
\begin{displaymath} 
{ \frac{{\bf K^-}}{{\bf \pi^-}}}
\end{displaymath}
\end{minipage}
\begin{minipage}[h]{10cm}
\epsfig{file=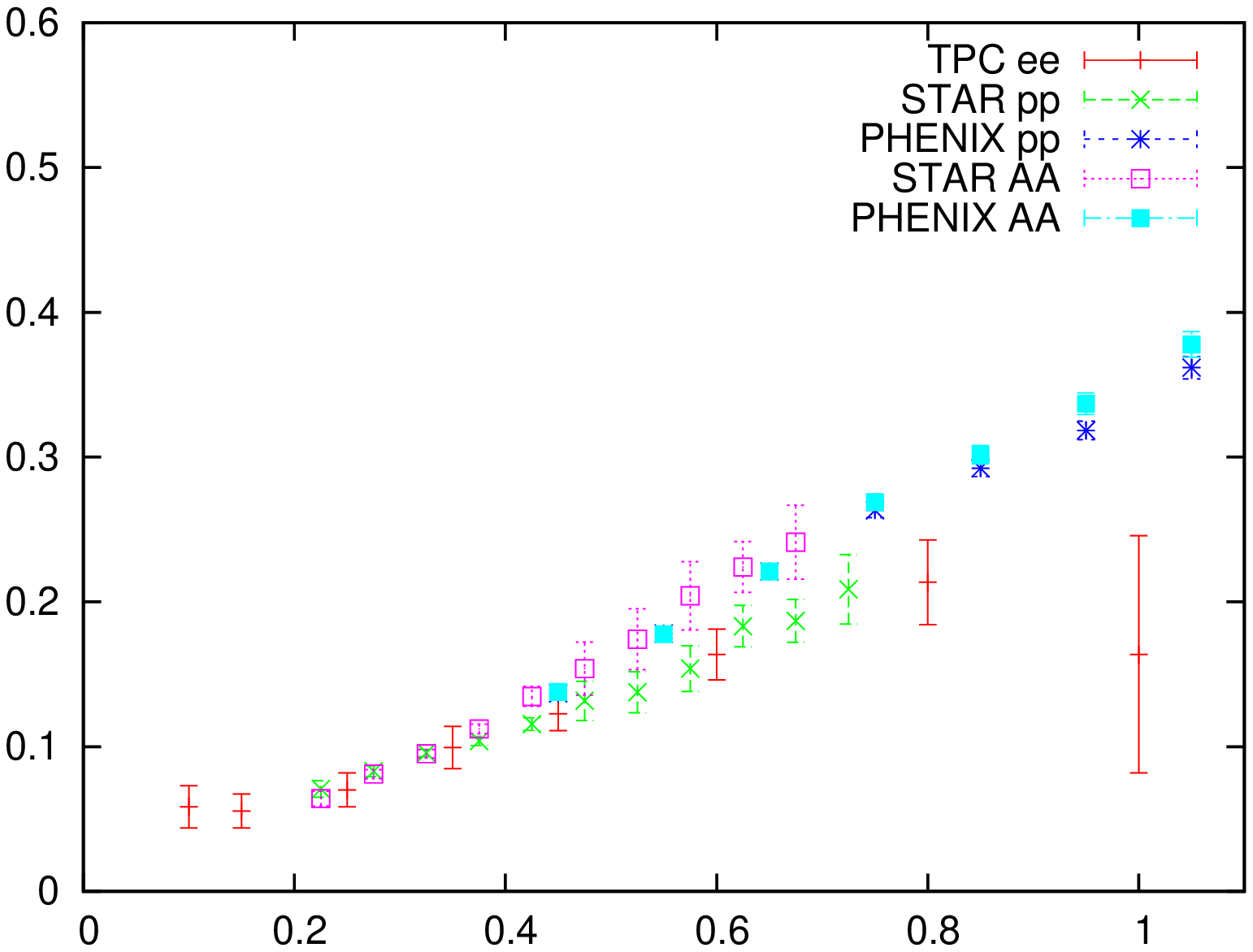,angle=-90,width=10.cm}
\hspace*{7cm} {\bf ${\bf p_T}$ [GeV]}
\end{minipage}
\end{center}
\caption{Convergence of particle ratios $K^-/\pi^-$ towards small $p_T$ for 
various processes: $e^+e^-$ annihilation 
(TPC \cite{tpc} data, $p_T$ with respect to sphericity axis), 
$pp$ (minimum bias) and central (0-5\%) $AuAu$
collisions (PHENIX \cite{phenix1} and STAR \cite{star1}
Collaborations).}
\label{fig:ratio}
\end{figure}
 
Finally, we may ask, whether the universal production of the soft particles
from gluon bremsstrahlung
also reflects in their composition as detected by particle ratios.
Such a universality can be expected if the source of the bremsstrahlung are
the colour triplet dipoles generated in $pp$ and $AA$ collisions by gluon
exchange. In that case not only hadronic collisions but also \epem
annihilations have universal dipole sources.

A similarity of particle ratios $K/\pi$ and $\bar p/\pi$ in the
$e^+e^-$ and $pp$ reactions at $p_T<0.5 $ GeV
has been indeed noted already some time ago by the TPC collaboration
\cite{tpc}. In this measurement the transverse momentum $p_T$ for the
particle
collisions was defined with respect to the sphericity axis.

The $p_T$ dependence of the particle ratios for several hadronic collisions
have been compared by PHENIX \cite{phenix1}.  While at the large $p_T>2$ GeV
the ratios $p/\pi$ and $K/\pi$ tend to approach large values $\sim 1$ in the
central $AuAu$ collisions, these ratios are reduced for non-central and
minimum bias $pp$ collisions.  Remarkably, these ratios converge for all the
different processes towards lower $p_T<1$ GeV.  In Fig.  \ref{fig:ratio} we
collect data in the low $p_T$ region on the ratio $K^-/\pi^-$ from the
$e^+e^-$, $pp$ and $AA$ interactions.  As one can see, these particle ratios,
indeed, approach each other towards low $p_T<0.4$~GeV pointing towards a
dominance
of multiple $q\bar q$ dipole radiation in all processes.

\section{Summary}
We note some universal features of the particle production in the limit
$p,p_T\to 0$ which we derive from the dominance of coherent 
QCD gluon bremsstrahlung  
in this limit. We consider the particle density $I_0$ in this limit for
which we predict 

1) the energy independence, and

2) the dependence on colour factors according to the minimal partonic
process (Born-term): 

a) \epem annihilation: $I_0^{g-jet}/I_0^{q-jet} = C_A/C_F$;

b) $pp$ scattering: $I_0^{pp}/I_0^{e^+e^-}=C_A/C_F$;

c) $AA$ scattering: $I_0^{AA}/I_0^{pp} = (N_{part}/2)\ C_A/C_A$.

These expectations are well met by the data. In consequence,
the soft particles do not
follow a universal thermal behaviour independent of the initial state.

There is also some universality in the particle ratios which tend to
converge to those from $q\bar q$ dipoles. Soft hadrons in the central region 
are produced first. In $AA$ collisions these slow particles stay behind and
do not participate in the equilibration.

It will be interesting to study the soft limit at LHC energies. If there are
new incoherent sources, as expected in some models with multiple
interactions, the soft density $I_0$ could start rising with energy.

\section*{Acknowledgement}
I would like to thank Valery A. Khoze and Misha G. Ryskin for the
collaboration and exchange about the content of this presentation and
to Andrzej Bialas for the interesting discussions about ``soft physics''.

\end{document}